\documentstyle[12pt]{article}

\input mssymb

\newcommand{\uu}[1]{\verb!#1!\endgroup}     %  is not expanded until
\newcommand{\mb}[1]{\ifmmode#1\else{\ifinner#1\else\mbox{$#1$ }\fi}\fi}

\newcommand\ga{\mb{\gamma}}

\newcommand\et{\mb{\eta}}

\newcommand\ka{\mb{\kappa}}

\newcommand\muu{\mb{\mu}} % use completely illogical abbreviation for
\newcommand\nuu{\mb{\nu}} % two-letter Greek symbols

\newcommand\si{\mb{\sigma}}

\newcommand\ph{\mb{\phi}}

\newcommand\ch{\mb{\chi}}

\newcommand\Th{\mb{\Theta}}

\newcommand\Ps{\mb{\Psi}}

\newcommand\calG{\mb{{\cal G}}}
\newcommand\calH{\mb{{\cal H}}}

\newcommand\calL{\mb{{\cal L}}}

\newcommand\calO{\mb{{\cal O}}}

\newcommand{\eq}[1]{\begin{equation}#1\end{equation}}
\newcommand{\eqn}[1]{\begin{eqnarray}#1\end{eqnarray}}

\font\bigm=cmex10 scaled\magstep1
\def\bigint_#1^#2{\hbox{\raise16.8pt\hbox{\bigm\char'132}}_{#1}^{\;\;#2}}

\font\Bigm=cmex10 scaled\magstep2
\def\Bigint_#1^#2{\hbox{\raise19.4pt\hbox{\Bigm\char'132}}_{\!#1}^{\:\;\;#2}}

\font\biggm=cmex10 scaled\magstep3
\def\biggint_#1^#2{\hbox{\raise23.2pt\hbox{\biggm\char'132}}_{\!#1}^{\;\;\;#2}}

\font\Biggm=cmex10 scaled\magstep4
\def\Biggint_#1^#2{\hbox{
\raise28.4pt\hbox{\Biggm\char'132}}_{\!\!#1}^{\:\;\;\;#2}}

\font\BIGm=cmex10 scaled\magstep5
\def\BIGint_#1^#2{\hbox{\raise33.5pt
\hbox{\BIGm\char'132}}_{\!\!\!#1}^{\:\;\;\;\;#2}}

\font\normalm=cmex10
\def\normalint_#1^#2{\hbox{\raise15pt\hbox{\normalm\char'132}}_{#1}^{\;\;#2}}

\def\smallint_#1^#2{\hbox{\raise10pt\hbox{\bigm\char'122}}_{#1}^{\;#2}}

\newdimen\mycirclesize
\newcount\mycirclesiz

\def\mycircl#1{%
\mycirclesiz\mycirclesize%
\mycirc{\the\mycirclesiz}%
}
\def\mycirc#1{%
\special{" newpath 0 0 #1 65536 div 0 360 arc stroke }%
}
\newcommand{\cinf}[1]{\mb{C^{\infty}(#1)}}
\newcommand{\x}{\mb{\times}}

\newcommand{\R}{\mb{\Bbb R}}
\newcommand{\hst}{\hspace{10mm}}
\newcommand{\fa}{\forall}

\newcommand{\g}{\mb{{\bf g}}}

\newcommand{\h}{\mb{{\bf h}}}
\newcommand{\hs}{\mb{{\bf h}^*}}

\newcommand{\iso}{\simeq}

\newcommand{\beq}{\begin{equation}}
\newcommand{\eeq}{\end{equation}}
\newcommand{\nn}{\nonumber}
\newcommand{\crc}{\mb{{\scriptstyle\circ}}}
\newcommand{\bea}{\begin{eqnarray}}
\newcommand{\eea}{\end{eqnarray}}

\newcommand{\lx}{\mb{\ltimes}}

\newcommand{\hbm}{\mb{\hbar^{-1}}}
\newcommand{\bs}{\backslash}

\def\one{{1\kern -.60em 1}}

\newcommand{\ci}{\cite}
\newcommand{\sect}{\section}
\renewcommand{\ll}{\label}

\renewcommand\o{\mb{{\cal O_\mu}}}

\topmargin = - 0.5 cm
\jot
\begin{document}
\renewcommand{\thefootnote}{\fnsymbol{footnote}}
%\maketitle
\bibliographystyle{unsrt}
\begin{flushright}
DAMTP 94-45
\end{flushright}

\begin{center}
{\Large Geometric quantization on homogeneous spaces and the meaning
of `inequivalent' quantizations\\}
\vspace*{0.5cm}
{ M.\ A.\
Robson\footnote{E-mail:
M.A.Robson\symbol{64}amtp.cam.ac.uk.}\footnote{Supported
by an E.P.S.R.C.\ studentship.}\\ }
\vspace*{0.2cm}
{\small Department of Applied Mathematics and Theoretical
Physics,\\
University of Cambridge, Silver Street,\\
Cambridge, CB3 9EW, U.\ K.\ }
\end{center}

\begin{abstract}
Consideration of the geometric quantization of the phase space of a
particle in an external Yang-Mills field allows the results of the
Mackey-Isham quantization procedure for homogeneous configuration
spaces to be reinterpreted. In particular, a clear physical
interpretation of the `inequivalent' quantizations occurring in that
procedure is given.\\
\\
\parbox[t]{2cm}{\em Keywords:}
\parbox[t]{10.77cm}{\em geometric quantization, homogeneous spaces,
external gauge fields}
\end{abstract}
%\newpage
\sect{Introduction}
Isham \ci{Isham} has given a quantization procedure for a particle
moving on a homogeneous configuration space, which was based upon the
earlier work of Mackey \ci{Mac}. Essentially quantization, in this
approach, corresponds to finding irreducible unitary representations
of a certain semidirect product group. We find that these
representations occur via the geometric quantization of the phase
space of a particle in an external Yang-Mills field. The symplectic
formulation of such a phase space is described in section
\ref{symform}, while the results of geometric quantization, as applied
to such spaces, is given in section \ref{gqn}. Note that our results
are applicable for non-homogeneous configuration spaces and are thus
presented for the general case.  An outline of the Isham-Mackey
quantization procedure is given in section \ref{I}, while the
comparison of the results of this method with those due to the
geometric quantization approach is given in section \ref{comp}.

The mathematical techniques we use are standard symplectic geometry
and Hilbert space representation theory. Full details of the results
of the geometric quantization procedure given in section~\ref{gqn} may
be found in \ci{MAR}.

\sect{Symplectic formulation of the phase
space} \ll{symform} We consider the case of a spin 0 particle moving
on an arbitrary Riemannian manifold $Q$ in an external gauge field
with gauge group $H$ (which we assume is compact). Mathematically,
this is realised as a connection on a principal $H$-bundle over $Q$;
the Yang-Mills field strength is identified with the local form of the
curvature associated to the connection. Wong \ci{Wong} introduced
equations of motion for such a particle and it has been shown
\ci{Mont,Sn,Wein78} that these are equivalent to the various
mathematical formulations as given by the Kaluza-Klein formulation of
Kerner \ci{Kerner}, and those of Sternberg \ci{Stern} and Weinstein
\ci{Wein78}. We use the symplectic formulation as detailed in
\ci{Wein78,Mont}, since this is the most suited to the method of
geometric quantization.

The right action of $H$ on $N$ lifts in a natural way to a symplectic
right action on $T^*N$ with an equivariant momentum map
$J:T^*N\to\hs$, where \hs is the dual of the Lie algebra of $H$. The
motion of the particle is described by an $H$-invariant Hamiltonian on
$T^*N$ (the `unconstrained' phase space). The metric on $N$ being
induced by the connection and the metric on $Q$. The symmetry
associated with the action of $H$ means that Marsden-Weinstein
reduction can be used to identify the reduced phase space, $P_\mu$, of
the system. Explicitly, $P_\mu=J^{-1}(\mu)/H_\mu$ for some $\mu\in\hs$
\ci{MW}. Here $H_\mu\subset H$ is the isotropy group of the point
$\mu$; the value of \muu reflects which subset of $T^*N$ the particle
is constrained to move in. Alternatively, the reduced phase space may
be identified with $P_\o=J^{-1}(\o)/H\subset (T^*N)/H$, which is
symplectically diffeomorphic to $P_\mu$ \ci{Marle,KKS}. Here
$\o\subset\hs$ is the coadjoint orbit containing the point
$\mu\in\hs$. Physically, $\o$, in its entirety, represents the charge
of the particle. Thus, the possible reduced phase spaces of the
particle are in one-to-one correspondence with its charge. Indeed,
each $P_\o$ is in fact a symplectic leaf in the Poisson manifold
$(T^*N)/H$, which inherits a Poisson bracket from the canonical one on
$T^*N$.

The r\^ole of the connection is that it allows a factorization of
$T^*N$ to be obtained, viz.\ $T^*N\iso N^\#\x\hs$ \ci{Mont}. Here
$N^\#$ is the pullback bundle of $N\to Q$ via the projection map
$\pi:T^*Q\to Q$, and inherits a right action of $H$. In this
trivialization the momentum map has the particularly simple form
$J(\ka_n,\nu)=\nu$, which means that we can take $P_\mu=N^\#/H_\mu$
and $P_\o=N^\#\x_H\o$ (i.e., $N^\#\x\o$ quotiented by the equivalence
relation $(\ka_n,\nu)\sim (\ka_n h,Ad^*_h\nu)$). Note that
$N^\#/H=T^*Q$, so the connection allows a projection from $P_\o$ to
$T^*Q$ to be defined. Considering $P_\o$ as the reduced phase space,
then, as noted by Weinstein
\ci{Wein78}, until the connection is chosen the variables conjugate to
position on $Q$ are inherently intertwined with the `internal'
variables associated to $\o$. Physically, this means without a
connection we cannot separate the the particle's external momentum
from its own `position' and `momentum' which is associated with the
motion on the coadjoint orbit $\o$.

\sect{Geometric quantization}
\ll{gqn}
{\em Notation.} We will denote the projection map of a bundle $C$ with
base space $X$ by $\pi_{C\to X}$.

The method of geometric quantization requires a complex line bundle,
$B$ (the prequantum line bundle), over $P_\mu$ together with a
connection on $B$ with curvature $\hbm\si$, where \si is the
symplectic two-form on $P_\mu$. It is well known that not all
symplectic manifolds admit such a line bundle. We find that the bundle
$B$ exists provided that there is a representation, $\ch_\mu$, of
$H_\mu$ into U(1) with gradient $i\hbm\mu$ at $e\in H_\mu$. This is
the same as Kostant's \ci{Kostant} formulation of the integrality
condition for the quantization of the coadjoint orbit $\o$. Indeed, we
shall see that the quantization of $P_\mu$ and \o is closely
connected.

Once the prequantum line bundle has been constructed, the next step is
to choose a polarization of $P_\mu$. We construct a polarization on
$P_\o\iso P_\mu$ in the following manner. On $T^*N\iso N^\#\x\hs$ we
choose the vertical polarization; this restricted to $N^\#$ gives an
$H$-invariant distribution $P_0$ on $N^\#$. Recall that the group $H$
is  assumed to be compact, which implies that $H_\mu$ is connected
\ci{GS}. Hence the coadjoint orbit \o has a natural $H$-invariant
positive polarization $P^\o$ \ci{Wood}. The direct sum of the
polarization $P^\o$ on \o and the distribution $P_0$ on $N^\#$ gives a
new $H$-invariant distribution on $N^\#\x\o$ which projects down to
give a polarization $P$ of $P_\o=N^\#\x_H\o$.

Attention is then restricted to the polarized sections of
$B$. Crucially, we find that they can be realised as sections of the
vector bundle $E=N\x_H\calH_\mu$, where $(n,v)\sim
(nh,\pi_\mu(h^{-1})v)$ for $h\in H$. Here $\calH_\mu$ is the Hilbert space
that arises when the coadjoint orbit \o is quantized. Specifically, it
consists of functions $\ph:H\to \Bbb C$ which satisfy
\eq{\ph(h_\mu h)=\ch_\mu(h_\mu)\ph(h)\hst\fa h_\mu\in H_\mu}
and are polarized with respect to the K\"ahler polarization on
$\o$. (Here the $\ph$ are regarded as defining a section of a line
bundle over $\o$.) An inner product on $\calH_\mu$ is given by
\eq{( \ph_1,\ph_2)_{\calH_\mu} =\int_{H_\mu\bs
H}d([h]_{H_\mu}) \langle \ph_1(h),\ph_2(h)\rangle _{\Bbb C}.}
The representation $\pi_\mu$ of $H$ is defined by
\eq{(\pi_\mu(h')\ph)(h)=\ph(hh').}
Further it is possible to show that the representation $\pi_\mu$ is
irreducible.

We identify the sections of $E$ with functions $\Ps:N\to\calH_\mu$
satisfying $\Ps(nh)=\pi_\mu(h^{-1})\Ps(n)$ for all $h\in H$. Let $\et$
be an $H$-invariant measure on $N$, which in turn determines a measure
$\nu$ on $Q$. We can define an inner product on the sections of $E$
via
\eq{(\Ps,\Ps')=\int_Q d\nu(\pi_{N\to Q}(n))\
(\Ps(n),\Ps'(n))_{\calH_\mu},}
and we restrict our attention to smooth functions \Ps that have
compact support. Let $C^\infty_c(Q)$ denote the subspace of smooth
functions on $Q$ with compact support; then, following \ci{LandSD}, we
consider the induced unitary representation, $\pi^\mu$, of the
semidirect product group Aut~$N\lx C^\infty_c(Q)$ acting on the
$\Ps's$ which is defined by
\eq{(\pi^\mu(\ph,f)\Ps)(n)=\left(\frac{d\et(\ph^{-1}n)}{d\et(n)}\right)^{1/2}
e^{-i\hbm f\crc\pi_{N\to Q}(n)}\Ps(\ph^{-1}n).\ll{therepnew}}
Here $\ph\in {\rm Aut\ }N$ is an element of the group of automorphisms
of $N$, i.e., diffeomorphisms of $N$ satisfying $\ph(n)h=\ph(nh)$. The
group law on Aut~$N\lx C^\infty_c(Q)$ being
$(\ph_1,f_1)\cdot(\ph_2,f_2)=(\ph_1\crc\ph_2,f_1+f_2\crc\bar\ph_1^{-1})$,
where $\bar\ph\in {\rm Diff\ }Q$ denotes the projection of $\ph$ to $Q$.
The representation $\pi^\mu$ is irreducible provided $N$ does not
decompose into a disjoint union of two Aut~$N$-invariant subsets both
of which have positive \et measure. Note that in the case $N=G$, a Lie
group, the [left] action of $G\subset{\rm Aut\ }G$ on $G$ is
transitive and thus this condition is then automatically satisfied.

The action of the generators of $\pi^\mu$ are, in fact, the quantum
operators corresponding to certain classical observables on
$P_\o$. The correspondence is given via a momentum map, $J_\mu$, for
the left action of Aut~$N\lx C^\infty_c(Q)$ on $P_\o$. (This action is
the standard one \ci{GS} of Diff~$N\lx\cinf N$ on $T^*N$ restricted to
Aut~$N\lx C^\infty_c(Q)\subset {\rm Diff\ }N\lx\cinf N$, then
restricted to acting on $J^{-1}(\o)$ and then dropped to give an action on
$P_\o=J^{-1}(\o)/H$.) Denoting the Lie algebra of a group \calG by
$\calL(\calG)$, then
the momentum map $J_\mu:P_\o\to\calL({\rm Aut\
}N\lx C^\infty_c(Q))^*$ is defined by
\eq{\langle J_\mu[p_n]_H, (A,f)\rangle=\langle p_n,A
\rangle+\pi^*_{N\to Q}f(n).}
Here $p_n\in J^{-1}(\o)\subset T^*N$ with $\pi_{T^*N\to N}(p_n)=n$,
while $A$ is a [right] $H$-invariant vector field on $N$ regarded as an
element in $\calL({\rm Aut\ }N)$ and $f\in C^{\infty}_c(Q)$, where the Lie
algebra of $C^{\infty}_c(Q)$ has been identified with the group itself.

We can use the map $J_\mu$ to define a map $\hat J_\mu:\calL({\rm Aut\
}N\lx\cinf Q)\to\cinf{P_\o}$ via $(\hat J_\mu(A,f))[p_n]_H=\langle
J_\mu[p_n]_H,(A,f)\rangle$ and it is this map which links observables
on $P_\o$ to generators of $\pi^\mu$. Following the convention that
the derived Lie algebra representation $d\tilde\pi$ of a
representation $\tilde\pi$ is given by
\eq{d\tilde\pi(A)=\left.i\frac{d}{dt}\tilde\pi(e^{tA})\right|_{t=0},}
then, we find that the quantum operator corresponding to the classical
observable $\hat J_\mu(A,f)$ is given by $\hbar d\pi^\mu(A,f)$. We
express this in terms of a `quantizing' map, $Q_\hbar$, which relates
classical observables to quantum operators. We write \eq{Q_\hbar(\hat
J_\mu(A,f))=\hbar d\pi^\mu(A,f).}  The operator $d\pi^\mu(A,f)$ acts
on, and is essentially self-adjoint on, compactly supported sections
of the vector bundle $E=N\x_H\calH_\mu$.

The expression for $\hat J_\mu$ is slightly more illuminating if we
use local coordinates for $P_\o=N^\#\x_H\o$.  Now $N^\#/H=T^*Q$, and
locally $P_{\calO_\mu}$ is like $(N^\#/H)\x\calO_\mu$. Thus, let
$(h^1,\ldots,h^{d_H},q^{d_H+1},\ldots,q^{d_N})$ be local coordinates
on $N$, where $(q^{d_H+1},\ldots,q^{d_N})$ are coordinates on $Q$ and
$(h^1,\ldots,h^{d_H})$ are coordinates on the fibre $H$. Further, let
$(p_{d_H+1},\ldots,p_{d_N})$ be the corresponding components of
covectors on $T^*Q$. Then, locally, we can label a point
$[p_n]_H$ in $N^\#\x_H\calO_\mu$ by
$(q^{d_H+1},\ldots,q^{d_N},p_{d_H+1},\ldots,p_{d_N},\nu)$. Physically,
the $q$'s and $p$'s represent the particle's external position and
momentum respectively, while \nuu represents the internal `position' and
`momentum'. We find
\eqn{\hat
J_\mu(A,f)[p_n]_H&=&v^\ga(q^{d_H+1},\ldots,q^{d_N})p_\ga\,+\langle
\nu,X(q^{d_H+1},\ldots,q^{d_N})\rangle\nn\\
&&\mbox{} +\,f(q^{d_H+1},\ldots,q^{d_N}).\ll{genobservables}} Here
$v^\ga$ are the components of the vector field $A$ on $N$ projected to
$Q$ and $X\in\h$ is related to $A$ via the connection. The repeated
index \ga is summed from dim~$H+1$ to dim~$N$.  Equation
(\ref{genobservables}) gives the local form of the observables which we
can quantize via $Q_\hbar$. Further, if we require the quantization
of a classical observable to give an essentially self-adjoint
operator, then the image of $\hat J_\mu$ is indeed the complete set of
observables that can be geometrically quantized in the normal manner.

\sect{Homogeneous configuration spaces}
\ll{I}
When the bundle $N$ is a finite-dimensional Lie group $G$ (with
$H\subset G$) the configuration space $Q=G/H$ is homogeneous. Isham
\ci{Isham} has considered quantization on such configuration spaces
and in this section we briefly outline his scheme.

The core idea of Isham's approach, for a general configuration space
$Q$, is to find a Lie group \calG which has a transitive Hamiltonian
action on $T^*Q$ together with a momentum map
$J_I:T^*Q\to\calL(\calG)^*$. Quantization then corresponds to finding
irreducible unitary representations of $\calG$. For such a
representation, $\pi_I$, the quantum operator corresponding to the
observable $\hat J_I(X)\in\cinf{T^*Q}$ is given by $\hbar d\pi_I(X)$,
i.e., $Q_\hbar(\hat J_I(X))=\hbar d\pi_I(X)$. Here $X\in\calL(\calG)$
and $\hat J_I:\calL(\calG)\to\cinf{T^*Q}$ is defined in the same
manner as $\hat J_\mu$ in section~\ref{gqn}.

Isham argues that the group \calG to use is any finite-dimensional
subgroup of $\hbox{Diff } Q\lx\cinf{Q}/\R$ (where $\Bbb R$ denotes the
functions constant on $Q$) whose action is still transitive on
$T^*Q$. In particular, for the homogeneous configuration space $Q=G/H$
Isham further motivates and justifies the choice of $\calG=G\lx
V^*\subset \hbox{Diff } Q\lx\cinf{Q}/\R$, where $V$ is a vector space
which carries an almost faithful representation of $G$ and there is a
$G$-orbit in $V$ that is diffeomorphic to $G/H$.

All the irreducible unitary representations of \calG can be
constructed using Mackey theory; specifically, they are induced from
irreducible unitary representations of subgroups $H'\lx
V^*\subset\calG$ where $H'$ is such that $G/H'\iso\Th$, where \Th is a
$G$-orbit in $V$. Now, in general, there will be more than one such
irreducible unitary representation of \calG corresponding to each
orbit. This immediately raises the first of two key questions, namely,
which of these `inequivalent' representations should be chosen?
Secondly, in addition to the problem posed by the first question, what
is the relevance of the representations corresponding to orbits
$\Th\iso G/H'$ where $H'\not\simeq H$? We can answer these questions
by considering the circumstances in which the geometric quantization
approach generates such representations.

\sect{The comparison}
\ll{comp}
Taking $N=G$ we see that Isham's phase space $T^*Q\iso
G^\#/H=P_{\mu=o}\subset (T^*G)/H$. Further, Isham's momentum map
$J_I$ corresponds to the restriction of the momentum map $J_{\mu=0}$ of
section \ref{gqn} to $\calG\subset {\rm Aut\ }N\lx C^{\infty}_c(Q)$. This
illustrates the point that the geometric quantization approach
considers \calG as a subgroup of ${\rm Aut\ }N\lx C^{\infty}_c(Q)$ rather than
${\rm Diff\ }Q\lx C^{\infty}_c(Q)$.

We can split the irreducible unitary representations of \calG into two
classes, those which arise from consideration of a $G$-orbit $\Th\subset V$
where $\Th\iso G/H$ (the first class) and those from a $G$-orbit
$\Th'\subset V$ where $\Th'\iso G/H'$ with $H\not\simeq H'$ (the
second class). We can now compare Isham's approach with the geometric
quantization one.
Specifically, the representations $\pi^\mu$ we find are the same as
those in Isham's first class. (Here we are restricting $\pi^\mu$ to
$\calG$.) Crucially, however, each of our representations corresponds,
via $\mu$, to a different symplectic leaf, $P_\o$, in $(T^*G)/H$. Further,
each symplectic leaf has a different momentum map and thus each of the
different representations corresponds to a  different classical
system. In terms of the quantizing map $Q_\hbar$, Isham considers the
phase space $G^\#/H\iso T^*Q$ with
\eq{Q_\hbar(\hat J_{\mu=0}(A,u))=\hbar d\pi^\mu(A,u),}
where $(A,u)\in\calL(\calG)^*\iso\g\x V^*$.
Note that it is not clear which representation $\pi^\mu$ is to be
chosen on the right hand side. Whereas we have the phase space
$G^\#\x_H\calO_\mu$ with
\eq{Q_\hbar(\hat J_{\mu}(A,u))=\hbar d\pi^\mu(A,u).}
It is now clear that different representations of \calG correspond to
different physical systems. In fact, for a particle moving in a
Yang-Mills field, the different representations of \calG correspond to
the different possible charges that the particle could have.

The representations of $\calG$ in Isham's second class clearly
correspond to the quantizations of constrained systems which have $H'$
as the symmetry (gauge) group. In terms of a particle  in a
Yang-Mills field, these representations correspond to a particle on
the configuration space $G/H'$ where the internal charge couples
to the gauge group $H'$. Thus, they are unrelated to the original system.

In conclusion, we take issue with the common interpretation of the
Mackey-Isham quantization scheme as providing a number of inequivalent
quantizations of a given classical system (viz. $T^*(G/H)$). Rather, each
representation of the `canonical group' \calG (Isham), or of the
system of imprimitivity defined by $G/H$ (Mackey), has been found to be
the unique (geometric) quantization of a specific classical phase
space.

\sect*{Acknowledgements}
The author would like to greatly thank N.\ P.\ Landsman for not only
suggesting this line of research but also for his many helpful
comments and suggestions.

\end{document}